# Colloidal Assemblies of Oriented Maghemite Nanocrystals and their NMR Relaxometric Properties †

Athanasia Kostopoulou,[a] Sabareesh K. P. Velu,[b] Kalaivani Thangavel,[b] Francesco Orsini,[b] Konstantinos Brintakis,[a,c] Stylianos Psycharakis,[a,d] Anthi Ranella,[a] Lorenzo Bordonali,[e] Alexandros Lappas,*[a] Alessandro Lascialfari*[b]

[a] *Institute of Electronic Structure and Laser, Foundation for Research and Technology - Hellas, Vassilika Vouton, 71110 Heraklion, Greece. Fax: +30 2810 301305; Tel: +30 2810 391344; E-mail: lappas@iesl.forth.gr*

[b] *Dipartimento di Fisica, Università degli studi di Milano and INSTM, via Celoria 16, I-20133 Milano, Italy. Fax: +39 02 50317208; Tel: +39 02 50317383; E-mail: alessandro.lascialfari@unimi.it*

[c] *Department of Physics, Aristotle University of Thessaloniki, 54124 Thessaloniki, Greece.*

[d] *Department of Medicine, University of Crete, Voutes, 71003 Heraklion, Greece.*

[e] *Dipartimento di Fisica, Università degli studi di Pavia and INSTM, via Bassi 6, I-27100, Pavia, Italy*

**Abstract**

Elevated-temperature polyol-based colloidal-chemistry approach allows for the development of size-tunable (50 and 86 nm) assemblies of maghemite iso-oriented nanocrystals, with enhanced magnetization. $^1$H-Nuclear Magnetic Resonance (NMR) relaxometric experiments show that the ferrimagnetic cluster-like colloidal entities exhibit a remarkable enhancement (4 ÷ 5 times) in the transverse relaxivity, if compared to that of the superparamagnetic contrast agent Endorem®, over an extended frequency range (1-60 MHz). The marked increase of the transverse relaxivity $r_2$ at a clinical magnetic field strength (~1.41 T), which is 405.1 and 508.3 mM$^{-1}$ s$^{-1}$ for small and large assemblies respectively, allows to relate the observed response to the raised intra-aggregate magnetic material volume fraction. Furthermore, cell tests with murine fibroblast culture medium confirmed the cell viability in presence of the clusters. We discuss the NMR dispersion profiles on the basis of relaxivity models to highlight the magneto-structural characteristics of the materials for improved $T_2$-weighted magnetic resonance images.

**A Introduction**

Over the past decade, there has been a considerable progress in the synthesis of colloidal inorganic nanocrystals and in their exploitation for applications in various fields, straddling from electronics and optoelectronics[1] to nanomedicine and clinical practice[2]. An interesting development in this direction involves the synthesis of colloidal nanocrystals made of multiple subunits arranged in a controlled topological fashion[3, 4] or self-assembled in cluster-like structures (e.g. of γ-$Fe_2O_3$, $Fe_3O_4$, PbS, ZnO, $TiO_2$ etc)[5-9]. Their synthesis is either the result of direct specific interactions (e.g. van der Waals attractions, steric repulsions, attractive depletion or capillary forces, Coulomb forces etc.) of their discrete components, or of some external stimulus, such as light, magnetic or electric fields.[10] The coexistence within the same nanoscale entity of distinct material sections, directly interconnected through inorganic interfaces, or secondary structures communicating through strong interactions, enables multifunctionality. This is because coupling between mechanisms is established across the interfaced material domains,[11,12,13] while the physical properties may also have a collective nature.[14] Based on these appealing features, such multifunctional systems are promising for technological areas entailing biology and biomedicine. Furthermore, those materials featuring a soft magnetic state in conjunction with some other non-homologous property (e.g. plasmonic or excitonic emission) and with appropriate surface biocompatible coatings, occupy a prominent position in this scenario: [4, 9] they can be exploited for *in-vivo* biomedical applications, such as magnetically driven contrast enhancement,[15, 16] selective hyperthermia treatment[17], and even targeted drug delivery[18].

Specifically, the afore-mentioned iron-oxides based nanomaterials emerge as promising probes in bionanotechnology as they allow for a novel approach to the diagnosis of various diseases with magnetic resonance imaging (MRI). During the last two decades, several commercial MRI contrast agents have been developed for clinical studies. The majority of such systems are based on superparamagnetic iron-oxide nanoparticles, coated with hydrophilic polymers or sugars, including Endorem®, Resovist®, and Combidex®. However, poly-dispersity and poor crystallinity are often the most common limitations of such systems, originating from the synthesis protocol; as a result, magnetic properties and contrast efficiencies vary from one batch to another. [19, 20]

To develop new magnetic compounds for improved MR-based diagnostics, it is desirable to adjust the magnitude of the particle's net magnetic moment µ (because $R_2 = 1/T_2 \propto \mu^2$, where $R_2$ is the transverse relaxation rate and $T_2$ is the spin-spin relaxation time) by controlling the structural characteristics of the nano-object, i.e. the nanoparticle's size, shape, composition and crystallinity.[21] Rapid advances in the chemical routes for the synthesis of size and shape controlled, surface functionalised magnetic nanoparticles,[22, 23] lead to increased particle magnetizations and enhanced image contrast, preserving biocompatibility at the same time. As individual Superparamagnetic Iron Oxide Nanocrystals (SPIONs) tend to give relatively low magnetization,[21] to improve the MRI contrast, larger colloidal entities composed of primary nanocrystals (PNCs), which are self-assembled[24] or incorporated into polymer matrices, have been investigated. The latter approach, for instance, entails SPIONs encapsulated in polyacrylamine[25], silica[26], phospholipid micelles[27], block polymers[28] or amphiphilic polymers[29, 30], while the former assembling approaches lead to efficient "clustering" of individual PNCs, and promote close contact. In turn, this pathway further boosts the magnetization of such nanostructures, while the proximity of nanoparticles within the cluster promotes the coupling of the neighboring magnetic moments and results in a stronger perturbation of the local magnetic field in their vicinity. Indeed, the hydrogen nuclear moments relax faster, leading to a strong contrast in the MRI images. Magnetic nanoclusters of this type have been developed with capping agents of different nature, including oleylamine/oleic acid,[31, 32] citrate,[33-35] polymers[7, 9, 24, 36-44] or block copolymers[45-47]. When the enlarged cluster-like entity contains a biocompatible polymer-coating such as polyvinyl alcohol (PVA), poly(ethylene glycol)- co fumarate (PEGF) or cross linked PEGF, wide-spread opportunities in biomedical field become possible.[24] These colloidal nanoclusters agents feature an improved contrast efficiency with respect to commercial compounds, or at least their MRI contrast response is comparable to that of the widely available compound Endorem®.

In the present work we report on size-controlled assemblies of pure maghemite ($\gamma$-Fe$_2$O$_3$) PNCs, which are crystallographically aligned within the hydrophilic colloidal nanoclusters (CNCs). Our system displays a weak ferrimagnetic behavior, in contrast to the majority of the known nanoparticle-based aggregates devised for MR-based applications. We suggest that careful engineering of the CNCs' magneto-structural characteristics produces a remarkable improvement of $r_2$ (about four-to-five times higher) compared to Endorem®. From the point of view of design criteria we suggest that an increased intra-aggregate magnetic material volume fraction and the proper choice of the surfactant, allowing the water proton penetration, is a necessary condition for the superior MR-related contrast enhancing features of such low-cytotoxicity nanoarchitectures.

## B Experimental

**B.1 Materials.**

All reagents were used as received without further purification. Anhydrous iron chloride (FeCl$_3$, 98%), was purchased from Alfa Aesar (United States). Anhydrous Sodium hydroxide (NaOH, 98%), polyacrylic acid (PAA, M$_w$= 1800), were purchased from Sigma Aldrich (United States), while Diethylene glycol (DEG, (HOCH$_2$CH$_2$)$_2$O) of Reagent (<0.3%) and Laboratory (<0.5%) grades were purchased from Fisher (United States). The absolute Ethanol was purchased from Sigma Aldrich.

**Cell cultures.**

NIH/3T3 murine fibroblasts (ATCC- American Type Culture Collection) were suspended at a concentration of 10$^5$ cells/mL in Dulbecco's modified Eagle's medium (DMEM) supplemented with 10% fetal bovine serum (FBS) and 1% antibiotic solution (GIBCO, Invitrogen, Karlsruhe, Germany); cells were then cultured at 37 °C, in an atmosphere of 5% CO$_2$, and detached when growth reached 75% confluence using 0.05% trypsin/EDTA (GIBCO, Invitrogen, Karlsruhe, Germany).

**B.2 Synthesis of γ-Fe$_2$O$_3$ Nanoparticle Clusters.**

The CNCs were synthesized by a modified high-temperature colloidal chemistry protocol based on a one-step polyol process first reported by Yin and co-workers.[48] All syntheses were carried out under argon atmosphere in 100-mL round-bottom three-neck flasks connected via a reflux condenser to standard Schlenk line setup, equipped with immersion temperature probes and digitally-controlled heating mantles. The reactants, FeCl$_3$, NaOH, polyacrylic acid (PAA, M$_w$= 1800), diethylene glycol (DEG), except Ethanol, were stored and handled under anaerobic conditions in an Ar-filled glove-box (MBRAUN, UNILab). Crucially, utilising two types of DEG grades allowed for assemblies of iron-oxide nanocrystals with small and large diameter to be prepared upon increasing water content in the reaction mixture. Observation of the purified nanocluster samples by transmission electron microscopy, at regular time-intervals over a period of a year, have shown that they are stable with no aggregation among their units. Details of the synthetic protocol can be found elsewhere.[49]

**B.3 Characterization techniques.**

**a) Transmission electron microscopy (TEM).**

Low-magnification and high-resolution TEM images were recorded on a LaB$_6$ JEOL 2100 electron microscope operating at an accelerating voltage of 200 kV. For the purposes of the TEM analysis, a drop of a diluted colloidal nanoparticle aqueous solution was deposited onto a carbon-coated copper TEM grid. All the images were recorded by the Gatan ORIUS$^{TM}$ SC 1000 CCD camera and the structural features of the nanostructures were studied by two-dimensional (2D) fast Fourier transform (FFT) analysis.

**b) Magnetic Measurements.**

The magnetic properties of the samples were studied by means of a Superconducting Quantum Interference Device (SQUID) magnetometer (Quantum Design MPMS XL5). The magnetic measurements were performed with the dried powder of the nanomaterials. The magnetic data have been normalized to the mass of the γ-Fe$_2$O$_3$, as derived by Inductively Coupled Plasma Atomic Emission Spectroscopy (ICP-AES) analysis ([Fe]$_{50.2\text{ nm CNCs}}$ = 44.9 ± 0.3 mM, [Fe]$_{85.6\text{ nm CNCs}}$ = 42.3 ± 0.3 mM).

**c) Nuclear Magnetic Resonance (NMR).**

Two solutions of CNCs were prepared for the NMR investigation, having an iron concentration C= 1.225 and 1.208 mmol/L, as determined by ICP-AES measurements for small and large CNCs, respectively. NMR longitudinal and transverse relaxation times were measured by using two different pulsed Fourier Transform (FT)-NMR spectrometers: i) a Smartracer Stelar relaxometer (using Fast-Field-Cycling, FFC, technique) for frequencies in the range 0.01≤ ν≤ 10 MHz, and ii) a Stelar Spinmaster relaxometer for ν> 10 MHz. We used a common saturation recovery sequence to measure the longitudinal (spin-lattice, $T_1$) nuclear relaxation time. Standard radio frequency excitation pulse sequences, CPMG for ν > 3 MHz and pre-polarized Hahn echo sequence for ν < 3 MHz, were used to measure the transverse (spin-spin, $T_2$) nuclear relaxation time. The investigated frequency range 0.01≤ ν≤ 60 MHz corresponds to an external magnetic field H= 0.00023 - 1.41 Tesla (ω= γ H, where ω is the Larmor frequency and γ/2π= 42.58 MHz/T). The magnetic field range was chosen in order to cover the typical fields used in both clinical and research laboratory MRI tomography (H= 0.2, 0.5 and 1.5 Tesla). The measurements at room and physiological temperatures gave the same results within 10 %.

The efficiency of the CNCs as MRI contrast agents was determined by calculating the nuclear longitudinal and transverse relaxivities, $r_1$ and $r_2$ respectively, defined as

$$r_i = \frac{(1/T_i)_{\text{meas}} - (1/T_i)_{\text{diam}}}{C}, \quad i = 1, 2 \qquad (1)$$

where $(1/T_i)_{\text{meas}}$ is the value measured for the solutions with concentration C (mmol/L) of the magnetic center and $(1/T_i)_{\text{diam}}$ represents the nuclear relaxation rate of the diamagnetic host solution, which in our case is water.

**d) Cell Viability Assay.**

The interaction of the cells with the CNCs was evaluated using the live-dead cell staining kit (Biovision). The assay protocol has been described in previous studies.[50, 51] Cell suspension ($10^5$ cells/mL) was deposited in 24-well plates and cultivated for 1, 3, 5 and 7 days. In parallel, different concentration of CNCs (25, 50, 100 and 200 μg/mL Fe) was added in the cell suspension for every different culture period. Cells that remained alive after certain days of cultivation were stained green whilst dead cells were stained yellow-red. The images were recorded in a Zeiss fluorescence microscope equipped with a Laser scanning system Radiance 2100 (400-700 nm) and a Carl-Zeiss Axio Camera HR. Experiments were repeated at least three times for each size and available concentration of the CNCs.

**e) Cell Proliferation Assay.**

Quantification of live cells after each day of culture was performed by using ATP-Glo ™ bioluminometric cell viability assay kit (Biotium, Inc.). This highly sensitive homogenous assay for quantifying Adenosine Triphosphate (ATP) involves an addition of ATP-Glo™ detection cocktail to cells cultured in a serum-supplemented medium. Since ATP is an indicator of metabolically active cells, the number of viable cells can be assessed based on the amount of ATP available. This ATP detection kit takes advantage of Firefly luciferase's use of ATP to oxidize D-Luciferin and the resulting production of light in order to assess the amount of ATP available. Luminescence measurements were performed using the Synergy™ HT Multi-Mode Microplate Reader (BioTek). The different CNCs concentrations were tested in duplicates in each individual cell culture experiment and these experiments were repeated at least three times. For statistical analysis, the data were subjected to one way ANOVA followed by Tukey tests for multiple comparisons between pairs of means, using commercially available software (SPSS 21, IBM).

**C Results and Discussion**

**C.1 Morphology and Crystal Structure**

Iron-oxide colloidal assemblies with two different sizes were prepared on the basis of a high-temperature wet chemistry polyol pathway. The influence of different synthetic parameters[49] on the structural characteristics of the CNCs were carefully examined in order to attain a significant degree of size/shape homogeneity. $^{57}$Fe Mössbauer spectra (MS) recorded for the two nanocrystal assemblies at 10, 77 and 300 K show similar features, confirming the maghemite ($\gamma$-$Fe_2O_3$) nature[49] of these iron-oxides. Furthermore, bright-field TEM images for two representative samples are shown in Figures 1a, d**.** Both CNCs samples have a flower-like, almost spherical shape without aggregation between units. Each nanocluster is an assembly of small PNCs, with no isolated nanoparticles left out of the aggregate. The average diameters of the entities are 50.2 ±5.4 and 85.6 ±13.3 nm (Figure 1a, c). Further evidence for the topological arrangement and the crystallographic orientation of the assembled nanocrystals is provided by HRTEM images and the calculated FFT patterns taken from individual CNCs (Figure 1b, d). The observed diffraction spots suggest that the individual nanocrystals are iso-oriented. Analysis of the FFT patterns allowed us to identify that the crystal symmetry of the CNCs as a whole is similar to that of the individual, small PNCs, which crystallize in the cubic spinel structure. The attractive magnetic dipolar interactions (vide infra) of the PNCs are strong enough to counter-balance the electrostatic repulsions and allow for the controlled assembly of the PNCs in a secondary structure. During the growth, the as-formed PNCs re-orient in the colloid and minimize their surface energy by aggregating through oriented attachment, and form a single-crystalline-like secondary nanostructure.[38]

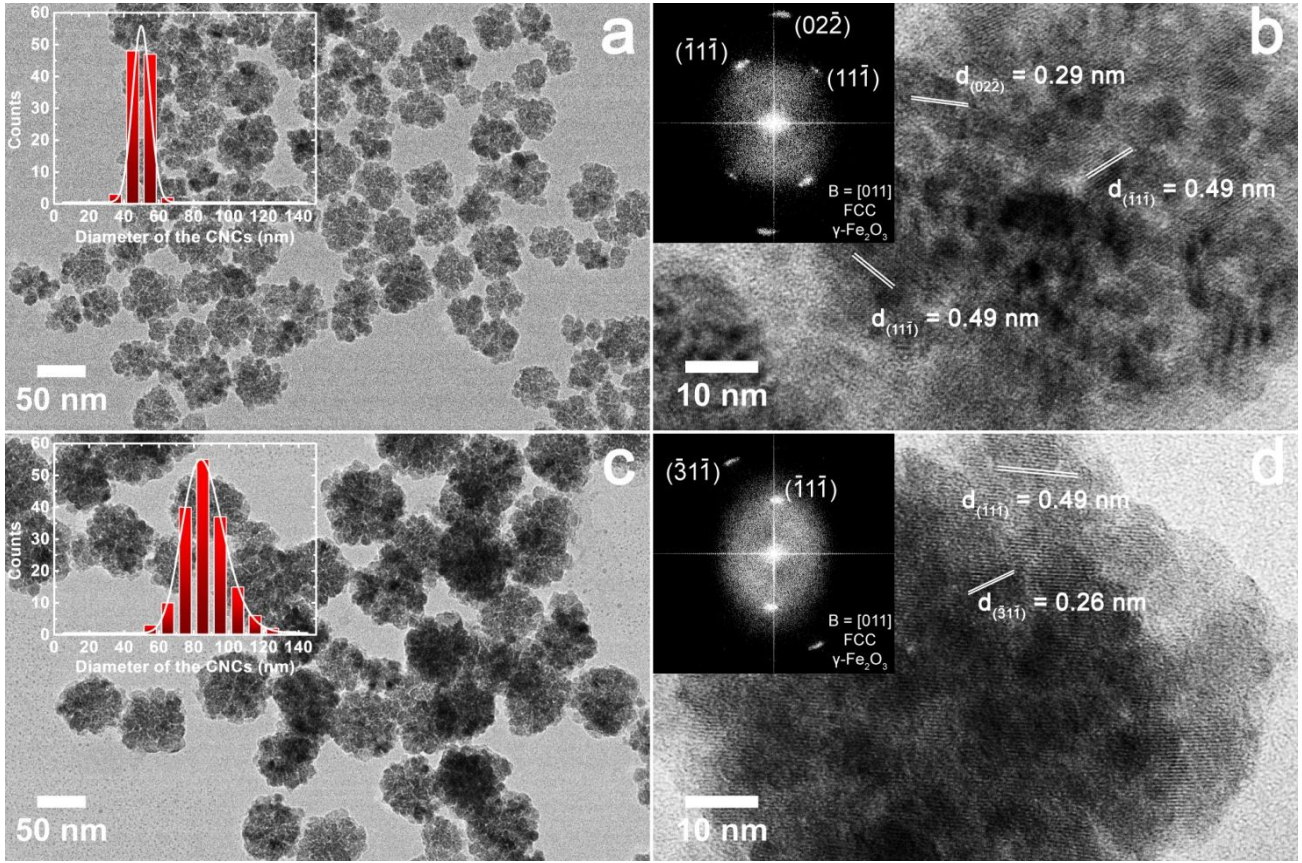

**Fig. 1** Low- (a, c) and high- (b, d) resolution TEM images of colloidal nanoclusters (CNCs) of maghemite. Insets to panels a, c show the size distribution for CNCs with diameter of 50 and 86 nm. Insets to panels b, d show the calculated FFT patterns from an isolated cluster-like structure; the reflections are indexed on the basis of a cubic spinel (fcc) iron-oxide crystal structure. The corresponding zone-axis is marked by B.

## C.2 Magnetic Properties

In Figures 2a,b the room temperature hysteresis loops indicate that the dried powders of the CNCs were ferrimagnetic (FiM) with coercive fields ($H_c$) of 5.4 and 2.1 Oe, for the small and the large CNCs, respectively, while the saturation magnetizations ($M_S$) were of 71.2 and 73.4 emu/g $\gamma$-$Fe_2O_3$, very close to the bulk value of 74 emu/g for $\gamma$-$Fe_2O_3$.

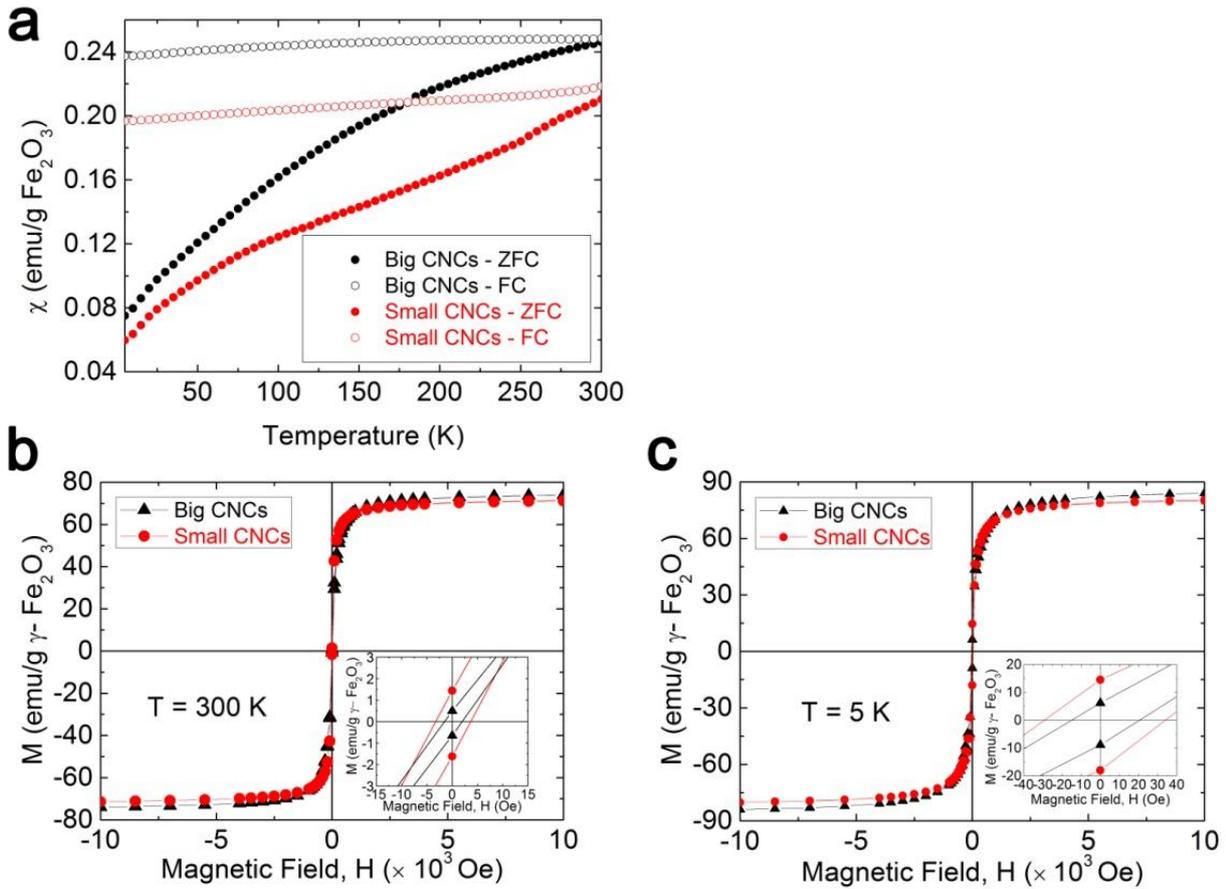

**Fig. 2** Magnetic properties of the small (50 nm) and large (86 nm) CNCs: (a) the temperature evolution of the ZFC and FC susceptibility, as well as hysteresis loops at 300 K (b) and 5 K (c).

The hysteretic behavior in the temperature dependent zero-field and field-cooled (ZFC-FC) susceptibility ($\chi$) indicates blocked particles below 300 K for both cases (Figure 2a). Since no obvious difference between the Mossbauer spectra of the dried powder and those of the frozen solutions of the CNCs could be resolved, we conclude that ferrimagnetism of the CNCs can be attributed to intra-cluster dipolar interactions only.[49]

**C.3 Cytotoxicity**

CNCs appear to satisfy the first prerequisite of tailored magnetic properties for improved MR properties. In addition, the purposeful choice of the PAA as surfactant renders the assemblies negatively charged (z-potential, was -65.4±10.8 and -50.0±6.5 mV for the small and the large CNCs, respectively) with good colloidal stability. However, a second condition for their implementation in the MR-based diagnostics requires verification of their toxicity on living cells. For this purpose, murine fibroblast cells (NIH/3T3) were exposed to the CNCs to thoroughly check the cell-nanocluster interactions. An ATP-based Glo$^{TM}$ cell viability assay was used for the luminometric measurement of the NIH/3T3 cells growth. Since the ATP indicates the metabolically active cells, we quantified the number of viable cells based on the concentration of the available ATP. ATP concentrations, recorded over a regular period of cultivation time (1, 3, 5 or 7 days), are plotted in Figures 3a, b. A series of different concentrations (25, 50, 100 and 200 μg/mL Fe) was used for both small and large CNCs.

Specifically, the cell growth for different concentrations of the CNCs, for the same number of days of culture, displayed no statistical difference when it was compared to the respective control samples (cells without CNCs). The significance levels (p) were estimated using one way ANOVA followed by Tukey tests (p<0.05). Additionally, the difference in the size of the CNCs did not affect the cell proliferation after 1, 3, 5 or 7 days of culture. The unaffected proliferation of NIH/3T3 cells under the presence of different concentrations of the CNCs provides valid evidence for the low toxicity of the present nanomaterial.

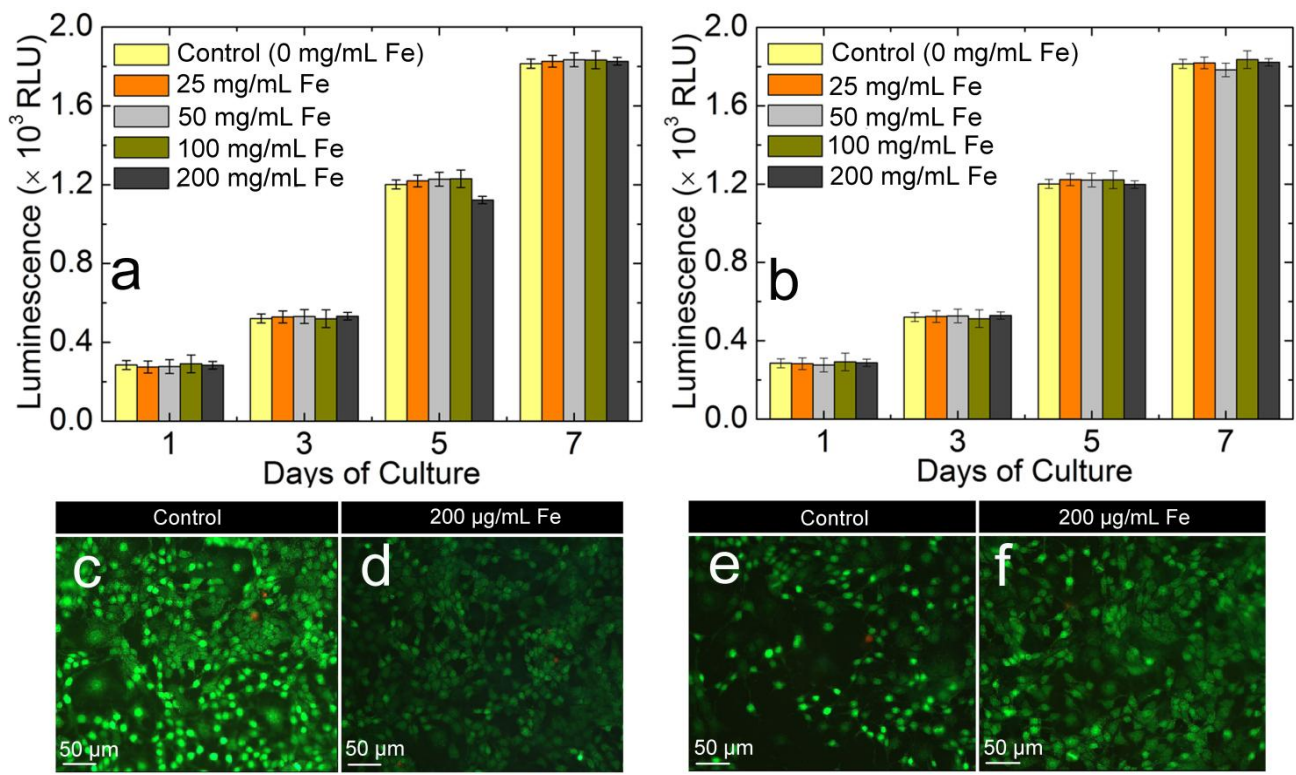

**Fig. 3** Cytotoxicity profile of the 50 (a) and 86 nm (b) CNCs samples. NIH/3T3 fibroblast cells were incubated with CNCs at the indicated doses for 1, 3, 5 and 7 days. ATP-Glo TM Bioluminometric cell viability assay kit was used for luminometric measurement of the fibroblasts' growth. Each histogram reflects the luminescence (mean value in RLU - relative luminescence units - ± Standard Deviation-SD) as derived from three independent experiments. Fluorescence microscopy images of live (green) and dead (orange-red) fibroblasts, cultured without CNCs (control) (c, e) and with small (d) or large (f) CNCs, after 5 days of culture.

The viability of fibroblasts was also tested by using a live-dead cell staining kit (BioVision). Cell viability was assessed after 1, 3, 5 and 7 days of culture with and without CNCs. Figures 3c-f show representative results from a 5-days cell culture and a concentration of 200 μg/mL Fe. The 3 and 7 days cultures, for all the CNCs concentrations (25-200 μg/mL Fe) show a similar behavior (Figure S1). In particular, these images depict the healthy fibroblasts (stained green), while dead cells (stained with PI, red) could only rarely be detected even on cell cultures with the highest measured concentration (200 μg/mL Fe) and the large CNCs (Figure 3f). In summary, fibroblast cells seem to tolerate the CNCs, showing high levels of viability, cell-cell interactions and low levels of toxicity, giving further support towards the potential use of the presently developed nanoarchitectures in MR-based diagnostics.

**C.4 The NMR Relaxation Properties and the Influence of the Intra-cluster Magnetic Material Volume Fraction.**

Let us now examine how the unique magnetostructural attributes of these nanocrystal assemblies can mediate the MR-based properties in view of their perspective imaging capability. The CNCs were shown to be ferrimagnetic, thus they are highly magnetized under an applied magnetic field and, consequently, they induce a substantial local perturbation of the dipolar magnetic field in their vicinity.[21] However, the possible increase of nuclear relaxation rates, giving a better efficiency in contrasting the MRI images, depends also on the dynamic electronic properties of the ensemble. To evaluate the physical properties of the system that influence the nuclear relaxation, we performed proton NMR relaxometry measurements. The longitudinal $r_1$ and transverse $r_2$ relaxivities (eq. (1)), were studied as a function of the measuring frequency for the two CNCs samples and compared with those obtained for Endorem® (Figure 4). Irrespective of the CNCs' size, the longitudinal $r_1$ relaxivity (~200 mM$^{-1}$ s$^{-1}$) is higher than that of Endorem® (10 mM$^{-1}$ s$^{-1}$) at low frequencies (10 kHz - 9 MHz), but falls a bit below the commercial product's values at higher frequencies (10 MHz - 60 MHz) (Figure 4a). The frequency behavior of $r_1$ is a consequence of the physical mechanisms responsible for the nuclear relaxation.[19, 52] In the low frequency (<1 MHz) range, the very high values of $r_1(\nu)$ and its flattening reflect the dominant role of the ferrimagnetic CNCs' high magnetic anisotropy. The absence of a maximum in $r_1(\nu)$, at $\nu > 1$ MHz, suggests that the diffusion process in this frequency range is poorly effective in shortening the nuclear relaxation of water protons (see comparison with Endorem, Figure 4a).

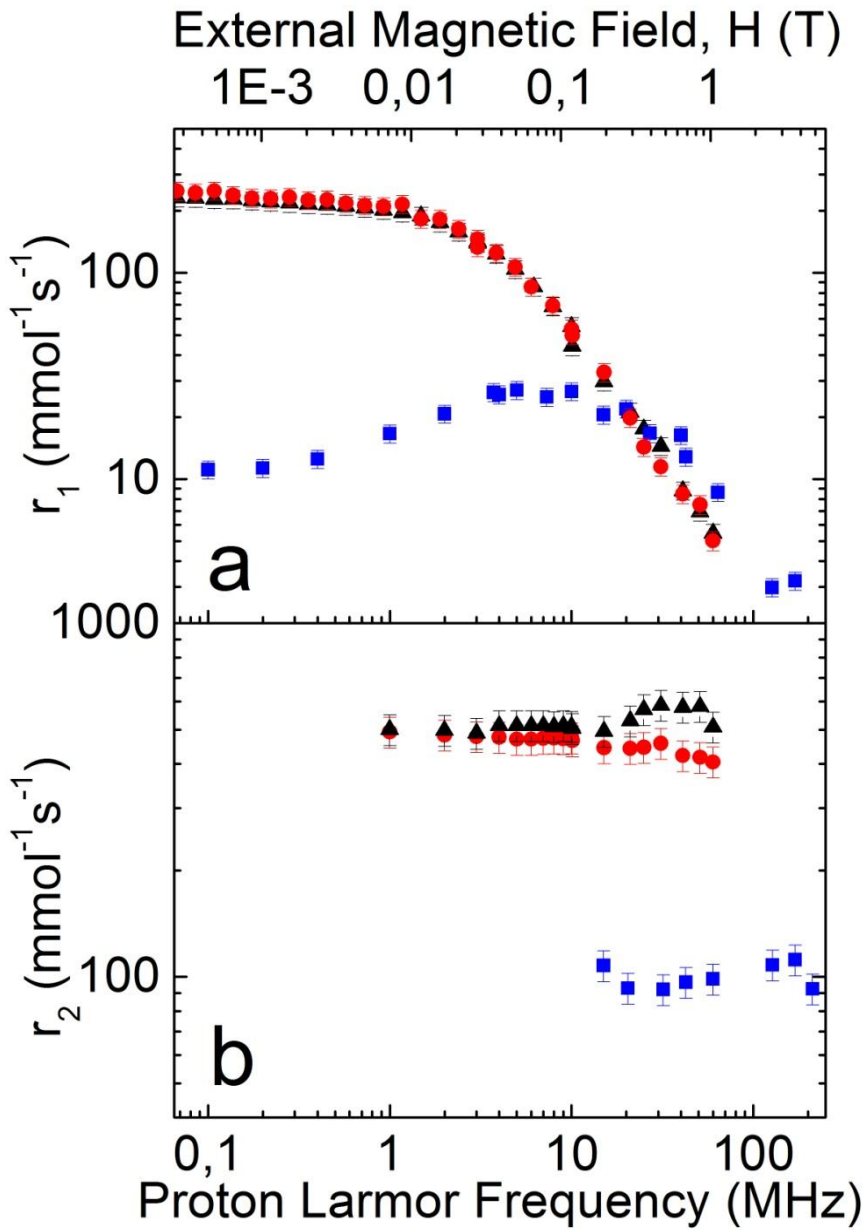

**Fig. 4** Room temperature longitudinal $r_1$ (a) and transverse $r_2$ (b) relaxivities as a function of proton Larmor frequency (or the external magnetic field, H) for the 50 (circles) and 86 nm (triangles) CNCs. The corresponding data for the until recently commercial contrast agent Endorem® (squares) are also shown.

On the other hand, the transverse $r_2$ relaxivity, i.e. the crucial relaxivity in $T_2$-relaxing materials, is enhanced by a factor ≈4 (≈5 for large CNCs) with respect to Endorem® (where $r_2$ ~ 100 mM$^{-1}$ s$^{-1}$), over almost all the investigated frequency range (Figure 4b). It is worth considering that a recent scaling-law approach has shown that the transverse relaxivity depends on three parameters, namely, the hydrodynamic diameter, the magnetization of the whole nanoparticle aggregate and the volume fraction of the magnetic material.[53] In that respect the $r_2$ values may be predicted by models assuming that the spin-spin relaxation occurs either in the motional averaging regime (MAR), where $\Delta\omega\cdot\tau_D < 1$ or in the static dephasing regime (SDR), where $\Delta\omega\cdot\tau_D > 5$; within the first regime the transverse relaxivity increases with the size, while in the second it remains almost unchanged.[54] To calculate the expected Larmor frequency shift we first estimated the values of the intra-cluster volume fraction, $\varphi_{intra}$, resorting to the following relation (refer to S2):

$$\rho_{\gamma\text{-Fe2O3}} (1 - f_{m\gamma\text{-Fe2O3}}) / (\rho_{PAA} f_{m\gamma\text{-Fe2O3}}) = (1 - \varphi_{intra}) / \varphi_{intra} \quad (2)$$

where $\rho_{\gamma\text{-Fe2O3}} = 4900$ kg/m$^3$ is the density of bulk maghemite, $\rho_{PAA} = 1150$ kg/m$^3$ is the density of the polyacrylic acid, and $f_{m\gamma\text{-Fe2O3}}$ is the weight fraction of the iron-oxide in a known mass of dried nanocluster powders, estimated from the thermogravimetric (TGA) measurements.[2] Eq. (2) yields $\varphi_{intra} = 0.60$ and $0.72$ for small and large CNCs, respectively.

On the other hand, the Larmor frequency shift is expressed as

$$\Delta\omega = \gamma\mu_0 M^*_V / 3 \quad (3)$$

where $M^*_V$ is the normalized magnetization (see Table I), while the translational diffusion time $\tau_D$ is defined as $\tau_D = d^2 / 4D$, where d is the diameter of the CNCs and $D = 3 \times 10^{-9}$ m$^2$/s is the water translational diffusion coefficient. $\tau_D$ amounts to ~0.2 and ~0.6 μs, for the small and large CNCs. Thus, finally:

$$\Delta\omega\cdot\tau_D = 4.9 \quad \text{and} \quad \Delta\omega\cdot\tau_D = 18.2 \quad (4)$$

for small (50 nm) and large (86 nm) CNCs respectively, indicating that the investigated samples fall at the extremes of the SDR regime ($5 < \Delta\omega\tau_D < 20$). We note that, within the SDR regime, the straddling water molecules feel a relatively constant dipolar magnetic field in their vicinity and it is not surprising that the experimentally observed $r_2(\nu)$ are marginally dependent on the CNCs' average size. The small difference in the sample volume magnetization, $M^*_V$, and $\varphi_{intra}$, is an additional reason behind the little variation at $\nu > 4$ MHz (Figure 4b).

The result of eq. (4) suggests that the transverse relaxivity ($r_2$) values for both CNCs samples may be predicted by models assuming that the spin-spin relaxation occurs in the SDR regime.

Firstly, if we calculate the expected $r_2$ value with the theoretical (approximate) expression found in ref. [53],

$$r_2^{theo} = 2\pi\gamma_P\mu_0 v_{mat} M^*_V / 9\sqrt{3}$$

where $v_{mat}$ is a volume fraction to iron concentration conversion factor ($v_{mat} = 1.57 \times 10^{-5}$ m$^3$mol$^{-1}$ for maghemite), and compare it to the experimental result $r_2^{exp}$, we observe a reasonable agreement, since $r_2^{theo}/r_2^{exp} = 1.11$ and 1.10. Conversely, the expected normalized relaxivity $r_2^* = r_2\, \varphi_{intra}/M^{*2}_V$ for 50 nm and 86 nm clusters is about $2\times10^{-8}$ and $8\times10^{-8}$ s$^{-1}$mM$^{-1}$m$^2$A$^{-2}$, respectively, but the real values characterizing the two samples are $5.5\times10^{-9}$ and $5.4\times10^{-9}$ s$^{-1}$mM$^{-1}$m$^2$A$^{-2}$, i.e. an order of magnitude different. Such a marked difference evidently hints at a poor agreement between theory and experiment; on the other hand, one may argue that samples located at the edges of the SDR regime are prone to misbehavior: the 50 nm sample is neither in the motional averaging regime (MAR) nor completely in the SDR range, i.e. in a region not described by any existing model; the 86 nm sample is very close to the $\Delta\omega\tau_D \sim 20$ limit, where the refocusing pulses used in the T$_2$ sequence become effective, an effect which is not accounted for by the SDR model. The experimental evidence proves, on the other hand, that Endorem® is not affected by these issues at all, since for the commercial compound $\Delta\omega\tau_D = 4.4$ and $r_2^* = 7.8\times10^{-8}$ s$^{-1}$mM$^{-1}$m$^2$A$^{-2}$, which is perfectly in line with the theory.

Finally, a justification for the significant (4 to 5 times) increase of $r_2$ with respect to the one of Endorem®, can be inferred by considering the almost two-times higher volume magnetization in conjunction with the roughly three-times larger intra-cluster volume fraction (Endorem®: $M^*_V = 0.77 \times 10^5$ A/m, $\varphi_{intra} = 0.23$; Table 1).

The above discussed points suggest that the $\varphi_{intra}$ parameter has a crucial role in the improvement of the MR properties.

**Table 1**. Parameters utilized for the calculation of the intra-cluster magnetic material volume fraction and the determination of the transverse relaxivity regime to which CNCs belong. These entail: $D_{hydro}$, the hydrodynamic diameter of the CNCs; $f_{m\ \gamma\text{-Fe2O3}}$, the weight fraction of the iron-oxide in a known mass of dried nanocluster powder as derived by the thermogravimetric (TGA) measurements; $\varphi_{intra}$, the intra-cluster volume fraction of the magnetic material; $M_S$, the saturation magnetization of the CNCs at room temperature; $M^*_V$, the volume magnetization; $\Delta\omega$, the Larmor frequency shift; $\tau_D$, the translational diffusion time. Note: parameter values for Endorem® tabulated in this table were taken from reference [53].

| Sample | $D_{hydro}$ (nm) | $f_{m\gamma\text{-Fe2O3}}$ (%) | $\varphi_{intra}$ | $M^*_V$ ($10^5$ A/m) | $\Delta\omega \cdot \tau_D$ | Regime |
|---|---|---|---|---|---|---|
| **Endorem®[1]** | 80 | 63.8 | 0.23 | 0.77 | 4.4 | SDR |
| **50.2 nm CNCs** | 78.6 | 86.8 | 0.60 | 1.95 | 4.9 | SDR |
| **85.6 nm CNCs** | 121.8 | 92.1 | 0.72 | 2.37 | 18.2 | SDR |

**C.5 The CNCs Relaxivities and their Relevance for other Ferrite Nanoachitectures**

Two broad categories, depending on the nature of their magnetic state, entailing either superparamagnetic (SPM) or ferrimagnetic (FiM) functional structures, are considered. A clear enhancement of the transverse relaxivity $r_2$ is presented against individual SPM nanocrystals with different capping agents such as, polymers [19], dendrons [55] or DHAA [56]. Their size-only dependent relaxivity in the MAR regime renders them less efficient $T_2$-relaxation agents compared to the CNCs, whose improved faster reduction of $T_2$ is determined by the synergetic action of $M_s$, size and $\varphi_{intra}$.[53] However, a progressive increase of the size of individual nanoparticles can lead to FiM nanoarchitectures which may provide enhanced relaxivites with the advantages of the SDR regime. The only FiM systems of large size and enhanced magnetic anisotropy are the iron-oxide nanocubes, with an edge length of 22 nm, encapsulated in PEG-phospholipids, with good colloidal stability; these systems have shown a value $r_2 = 761$ mM$^{-1}$s$^{-1}$, higher than our system [57] (see Table S2 for a more comprehensive comparison).

Alternatively, controlled aggregation of nanoparticles in secondary structures can further boost the relaxivities with an outcome equivalent to the SDR. In this respect, to the best of our knowledge, the only ferrimagnetic secondary structure that has been studied so far for its outstanding visualizing ($T_2$-weighted contrast properties) as well as drug-delivering actions, is a newly designed liposome-encapsulated magnetic nanoparticle cluster. [58] This system shows an unprecedented $r_2$ of 1286 mM$^{-1}$s$^{-1}$ but at a higher magnetic field of 2.35 T, thus leaving the CNCs a valid alternative.

Nevertheless, the strategy to embed nanoparticles in matrices/polymers [44,59] or aggregating them in well-controlled morphologies [32, 33, 36, 40, 45, 46, 60-62] can also afford SPM nanoarchitectures and shows relaxivities explainable within the SDR regime. In this case, our system shows larger $r_2$ (405 and 510 mM$^{-1}$s$^{-1}$ for the small and large assemblies, respectively) compared to most of the SPM cluster-type particles, such as those capped with citrate [33], PVP [36], polystyrene [63], Dextran [44], PEG [39], (Mal)mPEG-PLA copolymer [46], amine [32], TREG [62]. An exception is the SPM magnetite-based cluster-analogue of higher $r_2$ relaxivity, which is covered by the same surfactant as ours, but synthesized in an autoclave, through a polyol process for an extended period of time (12 hr). [60, 61] Amongst them, those clusters of diameter 34 and 63 nm (with $r_2$ of 540 and 630 mM$^{-1}$s$^{-1}$, correspondingly) have a $\varphi_{intra}$ of 0.30 and 0.50 and $M_V$ of 1.23 ×10$^5$ and 1.79 ×10$^5$ A/m, respectively, both smaller than the value for CNCs of the present study. The enhanced values of their relaxivities suggest that $r_2$ is not mediated only by the latter two magnetostructural parameters, but it must also be a function of the surface properties of the assemblies (e.g. thickness of surface coating, L, over the hydrodynamic diameter, $D_{hydro}$, of the assembly), as the interactions between the water protons and the assemblies could occur primarily on their surface. [64,65]

A further observation is that, in general, $r_2$ can be thought to originate from the variation of the diffusion length of the water molecules, relative to the dimensions of the surfactant-coordinated inorganic entities themselves. Then, the $r_2$ relaxivity is expected to decrease as the molecular weight of the capping agent increases [64], but this appears not to be the case of our system since the clusters are capped with a surfactant (PAA) of lower molecular weight ($M_w$= 1800 versus $M_w$= 5000[60]). The possible lower steric hindrance amongst shorter chains, allows for a higher packing density of a larger number of polyacrylate chains to coordinate the nanoclusters' surface. In turn, this complies with the higher z-potential attained by the CNCs in the present case (-65.4 and -50.0 mV for the small and the large clusters, compared to -38 and -43 mV, for 34 and 63 nm clusters by Li et al[60]). Such a dense surface coverage may impede the penetration ability (diffusivity) of the water molecules. As a consequence, the water-proton nuclear moments are less strongly perturbed by the local magnetic field generated by the inorganic entities at the surface of the nanoclusters themselves and generates a lower $r_2$.

Other differences may also be attributed to the likely modification of the chemical bonding (entailing the surface functionalised nanocrystals of the assembly) in the two cases. This bonding can mediate the particles' magnetic anisotropy or even enhance their surface spin disorder, thus having an impact in lowering the relaxivity values [66, 67].

**Conclusions**

The suggested colloidal chemistry pathway allows to obtain ferrimagnetic assemblies of crystallographically oriented primary nanocrystals of maghemite. They are not only well-dispersed in aqueous media and have low cytotoxicity, but show an enhanced transverse NMR relaxivity, $r_2$. The results suggest that the pronounced enhancements in the magnitude of $r_2$ have their origin in the ability to guide the assembly of individual nanocrystals so that they become crystallographically oriented, allowing for an increased magnetic material volume fraction ($\varphi_{intra}$) within the larger-grown nanocluster entities. The favourable high intra-cluster volume fraction of the magnetic moments appears to make them efficiently coupled to each other in the CNCs. In turn, this permits a coherent and intense perturbation of the dipolar magnetic field in the near vicinity of the straddling water molecules, a condition that enhances the relaxation of the associated proton nuclear spins. The polyacrylate surface coordinating groups of the CNCs further mediate the efficient penetration of the water protons and in conjunction with the surface spin disorder of the inorganic nanocrystals determine the magnitude of $r_2$. Such nanoarchitectures appear to have the potential for improved diagnostic quality in the $T_2$-weighted MR imaging techniques.

Finally, the observations presented in Section 3.4 lead to the conclusion that a definitive theory for $r_2$ relaxivity still has to be formulated, and a further effort needs to be sustained to account for overlooked effects that may indeed have a strong impact on the relaxometric properties of a nanoparticle-based contrast agent. Specifically, we would like to highlight the existence of classes of compounds that do not strictly obey the universal scaling law envisioned by Vuong et al[53] for clusters, since when the $\Delta\omega \cdot \tau_D$ is at lower and upper limits of the SDR range, predictions based on ref. [53] cannot be reliable.

**Acknowledgments**

This work was supported by the European Commission through the Marie-Curie Transfer of Knowledge program NANOTAIL (Grant no. MTKD-CT-2006-042459). SKPV, KVT, FO, LB and AL thank the Italian projects INSTM-Regione lombardia "Mag-NANO", FIRB "Riname"and Fondazione Cariplo n. 2010-0612. We thank Tomas Orlando and Paolo Arosio for experimental help.

**Notes and references**

1. D. V. Talapin, J. S. Lee, M. V. Kovalenko and E. V. Shevchenko, *Chem Rev*, 2010, **110**, 389-458.

2. P. Zrazhevskiy, M. Sena and X. Gao, *Chem. Soc. Rev.*, 2010, **39**, 4326-4354.

3. L. Carbone and P. D. Cozzoli, *Nano Today*, 2010, **5**, 449-493.

4. N. C. Bigall, W. J. Parak and D. Dorfs, *Nanotoday*, 2012, **7**, 282-296.

5. C. G. Li, Y. Zhao, F. F. Li, Z. Shi and S. H. Feng, *Chem Mater*, 2010, **22**, 1901-1907.

6. X. L. Hu, J. M. Gong, L. Z. Zhang and J. C. Yu, *Adv Mater*, 2008, **20**, 4845-7850.

7. X. L. Fang, C. Chen, M. S. Jin, Q. Kuang, Z. X. Xie, S. Y. Xie, R. B. Huang and L. S. Zheng, *J Mater Chem*, 2009, **19**, 6154-6160.

8. Y. Zhou and M. Antonietti, *J Am Chem Soc*, 2003, **125**, 14960-14961.

9. S. Xuan, Y. J. Wang, J. C. Yu and K. C. Leung, *Chem. Mater.*, 2009, **21**, 5079-5087.

10. M. Grzelczak, J. Vermant, E. M. Furst and L. M. Liz-Marzán, *Acs Nano*, 2010, **4**, 3591-3605.

11. K. H. Su, Q. H. Wei, X. Zhang, J. J. Mock, D. R. Smith and S. Schultz, *Nano Lett*, 2003, **3**, 1087-1090.

12. J. Lee, A. O. Govorov and N. A. Kotov, *Nano Lett*, 2005, **5**, 2063-2069.

13. A. Kostopoulou, F. Thétiot, I. Tsiaoussis, M. Androulidaki, P. D. Cozzoli and A. Lappas, *Chem. Mater.*, 2012, **24**, 2722-2732.

14. Z. Nie, A. Petukhova and E. Kumacheva, *Nature Nanotechnology*, 2010, **5**, 15-25.

15. H. B. Na, I. C. Song and T. Hyeon, *Adv Mater*, 2009, **21**, 2133-2148.

16. M. Lévy, C. Wilhelm, M. Devaud, P. Levitz and F. Gazeau, *Contrast Media Mol I*, 2012, **7**, 373-383.

17. C. S. S. R. Kumar, Mohammad, F., *Advanced Drug Delivery Reviews*, 2011, **63**, 789-808.

18. S. R. Deka, A. Quarta, R. Di Corato, A. Riedinger, R. Cingolani and T. Pellegrino, *Nanoscale*, 2011, **3**, 619-629.

19. M. F. Casula, P. Floris, C. Innocenti, A. Lascialfari, M. Marinone, M. Corti, R. A. Sperling, W. J. Parak and C. Sangregorio, *Chem Mater*, 2010, **22**, 1739-1748.

20. C. W. Jung and P. Jacobs, *Magn Reson Imaging*, 1995, **13**, 661-674.

21. J. Cheon and J. H. Lee, *Accounts Chem Res*, 2008, **41**, 1630-1640.

22. U. Jeong, X. W. Teng, Y. Wang, H. Yang and Y. N. Xia, *Adv Mater*, 2007, **19**, 33-60.

23. A. Figuerola, R. Di Corato, L. Manna and T. Pellegrino, *Pharmacol Res*, 2010, **62**, 126-143.

24. H. Amiri, M. Mahmoudi and A. Lascialfari, *Nanoscale*, 2011, **3**, 1022-1030.

25. B. A. Moffat, G. R. Reddy, P. McConville, D. E. Hall, T. L. Chenevert, R. R. Kopelman, M. Philbert, R. Weissleder, A. Rehemtulla and B. D. Ross, *Molecular Imaging*, 2003, **2**, 324-332.

26. E. Taboada, R. Solanas, E. Rodriguez, R. Weissleder and A. Roig, *Adv Funct Mater*, 2009, **19**, 2319-2324.

27. B. A. Larsen, M. A. Haag, N. J. Serkova, K. R. Shroyer and C. R. Stoldt, *Nanotechnology*, 2008, **19**, 265102.

28. J. F. Berret, N. Schonbeck, F. Gazeau, D. El Kharrat, O. Sandre, A. Vacher and M. Airiau, *J Am Chem Soc*, 2006, **128**, 1755-1761.

29. R. Di Corato, P. Piacenza, M. Musarò, R. Buonsanti, P. D. Cozzoli, M. Zambianchi, G. Barbarella, R. Cingolani, L. Manna and T. Pellegrino, *Macromol Biosci*, 2009, **9**, 952-958.

30. R. Di Corato, N. C. Bigall, A. Ragusa, D. Dorfs, A. Genovese, R. Marotta, L. Manna and T. Pellegrino, *Acs Nano*, 2011, **5**, 1109-1121.

31. T. Isojima, S. K. Suh, J. B. Vander Sande and T. Alan Hatton, *Langmuir*, 2009, **25**, 8292-8298.

32. K. C. Barick, M. Aslam, Y. P. Lin, D. Bahadur, P. V. Prasad and V. P. Dravid, *J Mater Chem*, 2009, **19**, 7023-7029.

33. L. Lartigue, P. Hugounenq, D. Alloyeau, S. P. Clarke, M. Lévy, J. C. Bacri, R. Bazzi, D. F. Brougham, C. Wilhelm and F. Gazeau, *Acs Nano*, 2012, **6**, 10935-10949.

34. F. Dong, W. Guo, J. H. Bae, S. H. Kim and C. S. Ha, *Chemistry- A European Journal*, 2011, **17**, 12802-12808.

35. C. Cheng, Y. Wen, X. Xu and H. Gu, *J Mater Chem*, 2009, **19**, 8782-8788.

36. S. Xuan, F. Wang, Y. X. Wang, J. C. Yu and K. C. F. Leung, *J Mater Chem*, 2010, **20**, 5086-5094.

37. M.

A. Daniele, M. L. Shaughnessy, R. Roeder, A. Childress, Y. P. Bandera and S. Foulger, *Acs Nano*, 2013, **7**, 203-213.

38. C. Cheng, F. Xu and H. Gu, *New J Chem*, 2011, **35**, 1072-1079.

39. F. Hu, K. W. MacRenaris, E. A. Waters, E. A. Schultz-Sikma, A. L. Eckermann and T. J. Meade, *Chem Commun*, 2010, **46**, 73-75.

40. J. Cha, Y. S. Kwon, T. J. Yoon and J. K. Lee, *Chem Commun*, 2013, **49**, 457-459.

41. S. M. Lai, J. K. Hsiao, H. P. Yu, C. W. Lu, C. C. Huang, M. J. Shieh and P. S. Lai, *J Mater Chem*, 2012, **22**, 15160-15167.

42. M. Mahmoudi, A. Simchi, M. Imani, A. S. Milani and P. Stroeve, *J Phys Chem B*, 2008, **112**, 14470-14481.

43. B. P. Jia and L. Gao, *J Phys Chem C*, 2008, **112**, 666-671.

44. E. K. Lim, E. Jang, B. Kim, J. Choi, K. Lee, J. S. Suh, Y. M. Huh and S. Haam, *J Mater Chem*, 2011, **21**, 12473-12478.

45. N. Pothayee, S. Balasubramaniam, N. Pothayee, N. Jain, N. Hu, Y. Lin, R. M. Davis, N. Sriranganathan, A. P. Koretsky and J. S. Riffle, *J Mater Chem B*, 2013, **1**, 1142-1149.

46. C. Zhang, X. Xie, S. Liang, M. Li, Y. Liu and H. Gu, *Nanomed-Nanotechnol*, 2012, **8**, 996-1006.

47. H. Chen, J. Yeh, L. Wang, H. Khurshid, N. Peng, A. Y. Wang and H. Mao, *Nano Res*, 2010, **3**, 852-862.

48. J. Ge, Y. Hu, M. Biasini, W. P. Beyermann and Y. Yin, *Angew Chem Int Edit*, 2007, **46**, 4342-4345.

49. A. Kostopoulou, K. Brintakis, M. Vasilakaki, K.N. Trohidou, A.P. Douvalis, A. Lascialfari, L. Manna, A. Lappas, *Nanoscale* 2014, **6**, 3764-3776.

50. A. Ranella, M. Barberoglou, S. Bakogianni, C. Fotakis and E. Stratakis, *Acta Biomater*, 2010, **6**, 2711-2720.

51. S. Psycharakis, A. Tosca, V. Melissinaki, A. Giakoumaki and A. Ranella, *Biomed Mater*, 2011, **6**, 045008.

52. S. Laurent, D. Forge, M. Port, A. Roch, C. Robic, L. V. Elst and R. N. Muller, *Chem Rev*, 2008, **108**, 2064-2110.

53. Q. L. Vuong, J. F. Berret, J. Fresnais, Y. Gossuin and O. Sandre, *Advanced Healthcare Materials*, 2012, **1**, 502-512.

54. E. Pöselt, H. Kloust, U. Tromsdorf, M. Janschel, C. Hahn, C. Maßlo and H. Weller, *Acs Nano*, 2012, **6**, 1619-1624.

55. B. Basly, D. Felder-Flesch, P. Perriat, C. Billotey, J. Taleb, G. Pourroy and S. Begin-Colin, *Chem Commun*, 2010, **46**, 985-987.

56. L. Xiao, J. Li, D. F. Brougham, E. K. Fox, N. Feliu, A. Bushmelev, A. Schmidt, N. Mertens, F. Kiessling, M. Valldor, B. Fadeel and S. Mathur, *Acs Nano*, 2011, **5**, 6315-6324.

57. N. Lee, Y. Choi, Y. Lee, M. Park, W. K. Moon, S. H. Choi and T. Hyeon, *Nano Lett*, 2012, **12**, 3127-3131.

58. G. Mikhaylov, U. Mikac, A. A. Magaeva, V. I. Itin, E. P. Naiden, I. Psakhye, L. Babes, T. Reinheckel, C. Peters, R. Zeiser, M. Bogyo, V. Turk, S. G. Psakhye, B. Turk and O. Vasiljeva, *Nature Nanotechnology*, 2011, **6**, 594-602.

59. M. S. Martina, J. P. Fortin, C. Menager, O. Clement, G. Barratt, C. Grabielle-Madelmont, F. Gazeau, V. Cabuil and S. Lesieur, *J Am Chem Soc*, 2005, **127**, 10676-10685.

60. M. Li, H. Gu and C. Zhang, *Nanoscale Res Lett*, 2012, **7**, 204.

61. F. Xu, C. Cheng, D. X. Chen and H. Gu, *Chemphyschem*, 2012, **13**, 336-341.

62. D. Maity, P. Chandrasekharan, P. Pradhan, K. H. Chuang, J. M. Xue, S. S. Feng and J. Ding, *J Mater Chem*, 2011, **21**, 14717-14724.

63. Y. T. Wang, F. H. Xu, C. Zhang, D. Lei, Y. Tang, H. Xu, Z. Zhang, H. Lu, X. Du and G. Y. Yang, *Nanomed-Nanotechnol*, 2011, **7**, 1009-1019.

64. S. Tong, S. J. Hou, Z. L. Zheng, J. Zhou and G. Bao, *Nano Lett*, 2010, **10**, 4607-4613.

65. D. Yoo, J. H. Lee, T. H. Shin and J. Cheon, *Accounts Chem Res*, 2011, **44**, 863-874.

66. L. Bordonali, T. Kalaivani, K. P. V. Sabareesh, C. Innocenti, E. Fantechi, C. Sangregorio, M. F. Casula, L. Lartigue, J. Larionova, Y. Guari, M. Corti, P. Arosio and A. Lascialfari, *J Phys-Condens Mat*, 2013, **25**.

67. H. Duan, M. Kuang, X. Wang, Y. A. Wang, H. Mao and S. Nie, *J Phys Chem C*, 2008, **112**, 8127-8131.

# Colloidal Assemblies of Oriented Maghemite Nanocrystals and their NMR Relaxometric Properties

**Athanasia Kostopoulou,**[a] **Sabareesh K. P. Velu,**[b] **Kalaivani Thangavel,**[b] **Francesco Orsini,**[b] **Konstantinos Brintakis,**[a,c] **Stylianos Psycharakis,**[a,d] **Anthi Ranella,**[a] **Lorenzo Bordonali,**[b] **Alexandros Lappas,\***[a] **Alessandro Lascialfari\***[b]

[a]Institute of Electronic Structure and Laser, Foundation for Research and Technology-Hellas, Vassilika Vouton, 71110 Heraklion, Greece
[b]Dipartimento di Fisica, Università degli studi di Milano and INSTM, via Celoria 16, I-20133 Milano, Italy
[c]Department of Physics, Aristotle University of Thessaloniki, 54124 Thessaloniki, Greece
[d]Department of Medicine, University of Crete, Voutes, 71003 Heraklion, Crete, Greece

**Corresponding Authors**
*lappas@iesl.forth.gr
*alessandro.lascialfari@unimi.it

## S1. Toxicity experiments

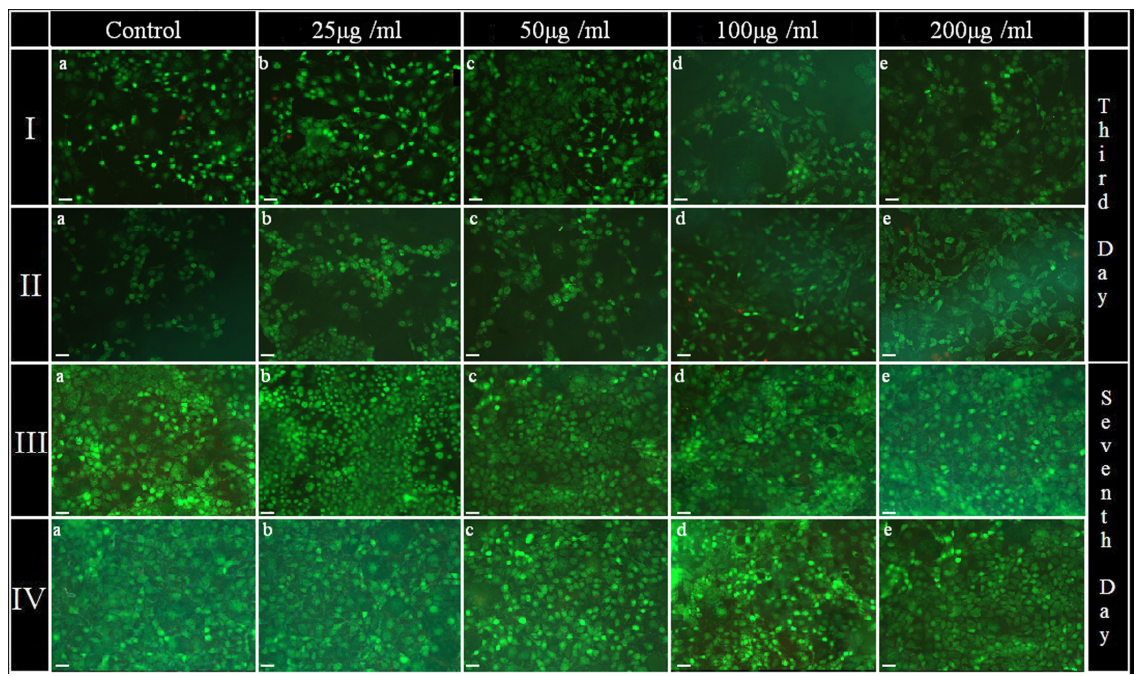

**Figure S1.** Fluorescence microscopy images of live (green) and dead (orange-red) fibroblast cells cultured without CNCs (a), with 85.6 nm CNCs (I, III), and 50.2 nm CNCs (II, IV) under different concentrations (25μg/mL-b, 50 μg/mL-c, 100 μg/mL-d, 200 μg/mL-e), after 3 (I,II) and 7 (III,IV) days of culture. The scale bar corresponds to 50 μm.

## S2. Calculation of the intra-cluster volume fraction of the magnetic material[1]

**Table S1.** (Copy of Table 1) Parameters utilized for the calculation of the intra-cluster magnetic material volume fraction and the determination of the transverse relaxivity regime where CNCs belong. These entail: $D_{hydro}$, the hydrodynamic diameter of the CNCs; $f_{m\ \gamma\text{-Fe2O3}}$, the weight fraction of the iron-oxide in a known mass of dried nanocluster powder as derived by the thermogravimetric (TGA) measurements; $\varphi_{intra}$, the intra-cluster volume fraction of the magnetic material; $M_S$, the saturation magnetization of the CNCs at room temperature; $M_V$, the volume magnetization; $\Delta\omega$, the Larmor frequency shift; $\tau_D$, the translational diffusion time.

| Sample | $D_{hydro}$ (nm) | $f_{m\ \gamma\text{-Fe2O3}}$ (%) | $\varphi_{intra}$ | $M_V$ ($10^5$ A/m) | $\Delta\omega\ \tau_D$ | Regime |
|---|---|---|---|---|---|---|
| **Endorem®[1]** | 80 | 63.8 | 0.23 | 0.77 | 4.4 | SDR |
| **50.2 nm CNCs** | 78.6 | 86.8 | 0.60 | 1.95 | 4.9 | SDR |
| **85.6 nm CNCs** | 121.8 | 92.1 | 0.72 | 2.37 | 18.2 | SDR |

The transverse relaxivity ($r_2$) regime in which the CNCs belong is determined mainly by three quantities, which are material-dependent characteristics, namely: the volume magnetization ($M_V$), the volume fraction of the magnetic material ($\varphi_{intra}$) and the hydrodynamic size (D). [1]

The intra-cluster volume fraction ($\varphi_{intra}$; Table S1) is calculated by the formula:

$$\frac{\rho_{\gamma-Fe_2O_3}(1-f_{m\ \gamma-Fe_2O_3})}{\rho_{PAA} f_{m\ \gamma-Fe_2O_3}} = \frac{1-\varphi_{intra}}{\varphi_{intra}} \quad (1)$$

where $\rho_{\gamma\text{-Fe2O3}}$ = 4900 kg/m$^3$, $f_{m\ \gamma\text{-Fe2O3}}$ the weight fraction of the iron-oxide in a known mass of dried nanocluster powders was estimated from the thermogravimetric (TGA) measurements and $\rho_{PAA}$= 1150 kg/m$^3$ the density of the polyacrylic acid.[2]

The normalized volume magnetization $M_V^*$, in A/m, can be calculated by the formula:

$$M_V^* = \varphi_{intra} \times M_S \times \rho_{\gamma-Fe_2O_3} \quad (2)$$

where the CNCs' room temperature saturation magnetizations are $M_S$= 71.2 and 73.4 emu/g for the 50.2 nm and 85.6 nm CNCs samples, respectively.

The Larmor frequency shift, $\Delta\omega$ is calculated by the formula:

$$\Delta\omega = \frac{\gamma\mu_0\ M_V^*}{3} \quad (3)$$

where $\gamma = 2.67513 \times 10^8$ rad s$^{-1}$T$^{-1}$ is the proton gyromagnetic ratio and $\mu_0 = 4\pi \times 10^{-7}$ T m A$^{-1}$ the magnetic permeability of the vacuum.

The estimated $\Delta\omega$ for the 50.2 nm and 85.6 nm CNCs samples are $2.37 \times 10^7$ s$^{-1}$ and $2.95 \times 10^7$ s$^{-1}$, respectively.

The translational diffusion time $\tau_D$ is given by the formula:

$$\tau_D = \frac{D_{hydro}^2}{4D} \qquad (4)$$

where $D_{hydro}$ is the hydrodynamic diameter of the CNCs, as calculated from the Dynamic Light Scattering (DLS) experiments and $D = 3 \times 10^{-9}$ m$^2$/s the water-proton translational diffusion time.

The calculated values of the $\tau_D$ are 0.2 and 0.6 μs for the 50.2 nm and 85.6 nm CNCs samples, respectively.

Based on the magnitude of the parameters derived from equations (3) and (4), ~5 < $\Delta\omega \tau_D$ < ~20 for the CNCs (Table S1). For this reason we are of the opinion that the transverse relaxivity (r$_2$) values for both CNCs samples can be predicted by models assuming that the spin-spin relaxation occurs in the static-dephasing-regime (SDR).

**Table S2.** Compilation of in-vitro transverse relaxivities, $r_2$, that allow comparison of the maghemite nanoclusters (CNCs) developed herein against other nanoarchitectures with potential in MR imaging. The materials were divided in two major categories depending on the nature of their magnetic state, namely, superparamagnetic (SPM) and ferrimagnetic (FiM). The regime, SDR or MAR, based on the outer sphere relaxation theory, in which they fall in is shown. The frequency, ν [= $(\gamma_\mu/2\pi)$ H], at which the relaxivity was measured, is also given.

| Magnetic State | Type of nanoarchitecture | Sample surface functionalisation | Regime | $r_2$ (mM$^{-1}$s$^{-1}$) | ν (MHz) |
|---|---|---|---|---|---|
| SPM | *Individual* | NCs capped with polymer [3] | MAR | 11 | 60 |
| | | NCs capped with dendrons [4] | MAR | 349 | 64 |
| | | NCs capped with DHAA [5] | MAR | 121 | 64 |
| | | Endorem[a] | SDR | 99 | 64 |
| | *Encapsulated in matrices* | NCs encapsulated in Dextran [6] | SDR | 312 | 64 |
| | | NCs encapsulated in liposomes[7] | SDR | 116 | 20 |
| | *Clusters* | Citrate-stabilized multi-core particles [8] | SDR | 365 | 9.25 |
| | | PVP-stabilized nanospheres [9] | SDR | 94 | 64 |
| | | polystyrene-capped nanoclusters [10] | SDR | 435 | 128 |
| | | Dextran-coated nanoclusters [6] | SDR | 312 | 64 |
| | | PEG-stabilized nanoflowers [11] | SDR | 238 | 64 |
| | | (Mal)mPEG-PLA copolymer-capped nanoclusters[12] | SDR | 465 | 60 |
| | | amine-stabilized nanoassemblies [13] | SDR | 315 | 85 |
| | | TREG-stabilized nanoclusters [14] | SDR | 295 | 400 |
| | | 34 nm and 63 nm PAA capped nanoclusters [15, 16] | SDR | 540 / 630 | 64 |
| FiM | *Individual* | nanocubes encapsulated in PEG-phospholipids our system [17] | SDR | 761 | 128 |
| | *Encapsulated in matrices* | liposome-encapsulated magnetic nanocluster [18]. | SDR | 1286 | 100 |
| | *Nanocrystal assemblies* [a] | PAA capped 50 nm nanoclusters[a] | SDR | 405 | 60 |
| | | PAA capped 86 nm nanoclusters[a] | SDR | 508 | 60 |

[a] results of the present work

1. Q. L. Vuong, J. F. Berret, J. Fresnais, Y. Gossuin and O. Sandre, *Advanced Healthcare Materials*, 2012, **1**, 502-512.
2. F. Xu, C. Cheng, F. Xu, C. Zhang, H. Xu, X. Xie, D. Yin and H. Gu, *Nanotechnology*, 2009, **20**, 405102.
3. M. F. Casula, P. Floris, C. Innocenti, A. Lascialfari, M. Marinone, M. Corti, R. A. Sperling, W. J. Parak and C. Sangregorio, *Chem Mater*, 2010, **22**, 1739-1748.
4. B. Basly, D. Felder-Flesch, P. Perriat, C. Billotey, J. Taleb, G. Pourroy and S. Begin-Colin, *Chem Commun*, 2010, **46**, 985-987.
5. L. Xiao, J. Li, D. F. Brougham, E. K. Fox, N. Feliu, A. Bushmelev, A. Schmidt, N. Mertens, F. Kiessling, M. Valldor, B. Fadeel and S. Mathur, *Acs Nano*, 2011, **5**, 6315-6324.
6. E. K. Lim, E. Jang, B. Kim, J. Choi, K. Lee, J. S. Suh, Y. M. Huh and S. Haam, *J Mater Chem*, 2011, **21**, 12473-12478.
7. M. S. Martina, J. P. Fortin, C. Menager, O. Clement, G. Barratt, C. Grabielle-Madelmont, F. Gazeau, V. Cabuil and S. Lesieur, *J Am Chem Soc*, 2005, **127**, 10676-10685.
8. L. Lartigue, P. Hugounenq, D. Alloyeau, S. P. Clarke, M. Lévy, J. C. Bacri, R. Bazzi, D. F. Brougham, C. Wilhelm and F. Gazeau, *Acs Nano*, 2012, **6**, 10935-10949.
9. S. Xuan, F. Wang, Y. X. Wang, J. C. Yu and K. C. F. Leung, *J Mater Chem*, 2010, **20**, 5086-5094.
10. Y. T. Wang, F. H. Xu, C. Zhang, D. Lei, Y. Tang, H. Xu, Z. Zhang, H. Lu, X. Du and G. Y. Yang, *Nanomed-Nanotechnol*, 2011, **7**, 1009-1019.
11. F. Hu, K. W. MacRenaris, E. A. Waters, E. A. Schultz-Sikma, A. L. Eckermann and T. J. Meade, *Chem Commun*, 2010, **46**, 73-75.
12. C. Zhang, X. Xie, S. Liang, M. Li, Y. Liu and H. Gu, *Nanomed-Nanotechnol*, 2012, **8**, 996-1006.
13. K. C. Barick, M. Aslam, Y. P. Lin, D. Bahadur, P. V. Prasad and V. P. Dravid, *J Mater Chem*, 2009, **19**, 7023-7029.
14. D. Maity, P. Chandrasekharan, P. Pradhan, K. H. Chuang, J. M. Xue, S. S. Feng and J. Ding, *J Mater Chem*, 2011, **21**, 14717-14724.
15. F. Xu, C. Cheng, D. X. Chen and H. Gu, *Chemphyschem*, 2012, **13**, 336-341.
16. M. Li, H. Gu and C. Zhang, *Nanoscale Res Lett*, 2012, **7**, 204.
17. N. Lee, Y. Choi, Y. Lee, M. Park, W. K. Moon, S. H. Choi and T. Hyeon, *Nano Lett*, 2012, **12**, 3127-3131.
18. G. Mikhaylov, U. Mikac, A. A. Magaeva, V. I. Itin, E. P. Naiden, I. Psakhye, L. Babes, T. Reinheckel, C. Peters, R. Zeiser, M. Bogyo, V. Turk, S. G. Psakhye, B. Turk and O. Vasiljeva, *Nature Nanotechnology*, 2011, **6**, 594-602.